\documentclass[aps, preprint, onecolumn, superscriptaddress, floatfix, nofootinbib]{revtex4-1}

\usepackage{graphicx}
\usepackage{bm}
\usepackage{amsmath}
\usepackage{amsfonts}
\usepackage{amssymb}
\usepackage{multirow}
\usepackage[colorlinks,linktocpage,linkcolor=cyan,citecolor=cyan]{hyperref}
\usepackage{adjustbox}
\usepackage{array}
\usepackage[capitalise]{cleveref}
\usepackage[utf8]{inputenc}

\begin{document}


\title{Stochastic gravitational wave background due to gravitational wave memory}

\author{Zhi-Chao Zhao}%
\affiliation{Department of Astronomy, Beijing Normal University, Beijing 100875, China}%

\author{Zhoujian Cao}%
\email{Correspondence author: zjcao@amt.ac.cn}
\affiliation{Department of Astronomy, Beijing Normal University, Beijing 100875, China}%
\affiliation{School of Fundamental Physics and Mathematical Sciences, Hangzhou Institute for Advanced Study, University of Chinese Academy of Sciences, Hangzhou 310024, China}%

\begin{abstract}
    Gravitational wave memory is an important prediction of general relativity, which has not been detected yet. Amounts of memory events can form a stochastic gravitational wave memory background. Here we find that memory background can be described as a Brownian motion in the condition that the observation time is longer than the averaged time interval between two successive memory events. We investigate, for the first time, the memory background of binary black hole coalescences. We only consider the spectrum of the memory background for a relatively low frequency range. So we can use the step function to approximate the waveform for each memory event. Then we find that the spectrum is a power law with index -2. And the amplitude of the power law spectrum depends on and only on the merger rate of the binary black holes. Consequently, the memory background not only provides a brand new means to detect gravitational wave memory but also opens a new window to explore the event rate of binary black hole mergers and the gravity theory. Space-based detectors are ideal to detect the gravitational wave memory background which corresponds to supermassive binary black holes. Since gravitational wave memory is only sensitive to the merger stage of binary black hole coalescence, the memory background will be an ideal probe of the famous final parsec problem.
\end{abstract}


\maketitle

\section{Introduction}\label{section1}
Gravitational waves (GW) from binary black hole mergers leave a permanent imprint on space-time, called gravitational wave memory \cite{Zeldovich74,Pay83,Braginsky:1986ia,braginsky1987gravitational,christodoulou1991nonlinear,Fra92}. GW memory is related to the asymptotic symmetry of spacetime and the soft theorem \cite{strominger2016gravitational,PhysRevD.95.125011,PhysRevLett.116.231301,pasterski2016new}. The detection of GW memory may reveal fundamental issues of physical laws \cite{du2016gravitational,hollands2017bms,PhysRevLett.121.071102,Hou2021Gravitational,2021JenkinsNonlinear,PhysRevD.103.104026} through the infrared triangle \cite{2017JHEP...09..154P}. The pulsar timing arrays (PTAs) may detect GW memory generated by the mergers of supermassive black hole binaries \cite{haasteren2010gravitational,CorJen12,MadCorCha14,2015MNRAS.446.1657W}. But the rates of detectable mergers are expected to be too low for practical detection \cite{Aggarwal2020The,PhysRevD.99.044045}. For LIGO-like detectors, the quasi-direct behavior of GW memory makes the detection of a single memory event extremely hard \cite{lasky2016detecting} (but see Ref.~\cite{2021arXiv211007754S} for special data analysis technique). Multiple events are consequently needed for GW memory detection \cite{lasky2016detecting}. Accurate waveform models \cite{favata2009post,favata2009nonlinear,favata2010gravitational,pollney2010gravitational,Cao16,mitman2021adding,liu2021accurate,PhysRevD.104.064056} may optimistically help the multiple events method realize the GW memory detection in the coming years \cite{PhysRevD.101.083026,hubner2020measuring,PhysRevD.104.023004}.

Even though there is no clear detection of GW memory, the existence of GW memory is a certain prediction of general relativity \cite{christodoulou1991nonlinear}. The nature of a single GW memory event has been extensively studied. But the superposition of multiple GW memory events has attracted much less attention before. A stochastic background of GW memory could be formed by composing multiple GW memory events \cite{PhysRevLett.118.181103}. Moreover, the stochastic background of GW memory may be detected even though the detection of a single event is impossible. For example the usual non-memory stochastic gravitational wave background [hereafter we call it non-memory SGWB to distinguish the stochastic gravitational wave memory background (SGWMB) which represents the memory part of SGWB] is much easier for PTA to detect than a single event of binary black hole (BBH) merger \cite{2020ApJ...905L..34A}.

Similar to the stochastic gravitational wave background generated by the non-memory gravitational wave part, the SGWMB can be analyzed through GW energy density \cite{PhysRevLett.118.181103}. The GW energy density analysis method is valid for those superpositions whose waveforms overlap. Due to the burst behavior of GW memory events, consecutive GW memory events may not overlap \cite{PhysRevResearch.2.012034}. Specifically, the GW memory events produced by BBH mergers typically do not overlap. In the following, we will show that the stochastic gravitational wave memory background composing non-overlapping GW memory events can be described by a Brownian motion. Based on our Brownian motion model, we find that the power spectrum index of SGWMB is -2 which is characteristically different to that of the non-memory GW part background of compact binary coalescence.

\section{From a single event of gravitational wave memory to a background}\label{sec:2}
Bursts of GWs, especially those from BBH mergers, cause a small permanent change of metric and leave a permanent imprint on space-time \cite{christodoulou1991nonlinear}. Based on Bondi-Metzner-Sachs (BMS) balance relation \cite{liu2021accurate,mitman2021adding} we can calculate the GW memory of a BBH merger event if only the binary parameters are known \cite{PhysRevD.104.064056}. Using spin-weighted $-2$ spherical harmonic functions $Y_{-2lm}$ to decompose the GW memory $h\equiv h_{+}-ih_{\times}$, we have
\begin{align}
&h(q,M,\vec{\chi}_1,\vec{\chi}_2,d_L,\iota,\phi_c)\equiv\nonumber\\
&\sum_{l=2}^{\infty}\sum_{m=-l}^lh_{lm}(q,M,\vec{\chi}_1,\vec{\chi}_2,d_L)Y_{-2lm}(\iota,\phi_c),
\end{align}
where $q$, $M$, $\vec{\chi_{1,2}}$, and $d_L$ are the mass ratio, the total mass, the spins, and the luminosity distance of the binary, $\iota$ is the inclination angle of the orbital plane, and $\phi_c$ is the gravitational wave phase at merger. In \cite{PhysRevD.104.064056} we have shown that the memory can be well approximated as
\begin{align}
&h(q,M,\vec{\chi}_1,\vec{\chi}_2,d_L,\iota,\phi_c)\approx h_{20}(q,M,\vec{\chi}_1,\vec{\chi}_2,d_L)Y_{-220}(\iota).
\end{align}

GW background means a lot of GW signals acting on the detector. We can express the SGWMB as
\begin{align}
\mathfrak{M} = \sum_{j=1}^{\infty}&\Re[(F^+(\theta_j,\phi_j,\psi_j)+iF^\times(\theta_j,\phi_j,\psi_j))\times\nonumber\\
&h(q_j,M_j,\vec{\chi}_{1j},\vec{\chi}_{2j},d_L,\iota_j,\phi_{cj})]\label{eq1}
\end{align}
where $j$ is the index denoting the $j$-th memory signal, and $F^{+,\times}$ are the pattern functions of the detector.

\section{Brownian motion model of SGWMB}\label{sec:3}
\begin{figure}
    \centering
    \includegraphics[width=1.0\columnwidth]{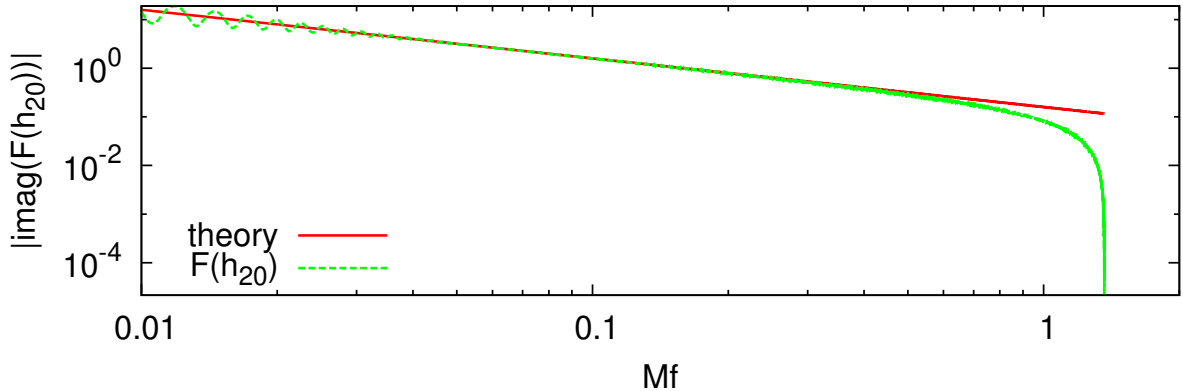}
    \caption{Fourier transform comparison between the step function and memory waveform. The red solid line corresponds to the Fourier transformation of step function shown in (\ref{eq6}). The green dashed line corresponds to the Fourier transformation of the scaled memory waveform. $M$ is the total mass of the BBH. $h_{20}$ means the spin weighted spherical harmonic mode (2,0).}
    \label{fig1}
\end{figure}

When GW memory relating to a binary black hole merger happens relatively fast with respect to an SGWMB observation, we can approximate such an individual memory signal as a step function of time \cite{CorJen12}. But we have to note the valid frequency range for such approximation. We have Fourier transformation of step function as
\begin{align}
\mathcal{F}[\Theta(t)]=\frac{1}{2}[\delta(f)-\frac{i}{\pi f}],\label{eq6}
\end{align}
where $\delta$ is the Dirac delta function. Taking a typical memory waveform of spinless BBH with equal mass and scaling it into the range $(0,1)$, we compare its Fourier transformation to the above result for step function in Fig.~\ref{fig1}. The oscillation of the line corresponding to the memory waveform at lower frequency side is due to the frequency leakage of fast Fourier transformation (FFT). Theoretically, we are sure the consistency between these two results is better when the frequency is lower. From this comparison, we can see the step function is a good approximation to the memory waveform for the frequency range $Mf\lesssim0.5$ where $M$ is the total mass of the binary black hole. This finding is consistent with the result shown in Fig.~3 of \cite{PhysRevLett.118.181103}.

Based on the step function approximation, the SGWMB (\ref{eq1}) behaves as a sum of a lot of step functions with respect to time. The amplitude and the sign of each step function are determined by the property of the corresponding binary \cite{PhysRevD.104.064056}. We treat the parameters of each binary merger as random variables which reduce a random amplitude and a random sign of the corresponding step function.
\begin{figure}
    \centering
    \includegraphics[width=1.0\columnwidth]{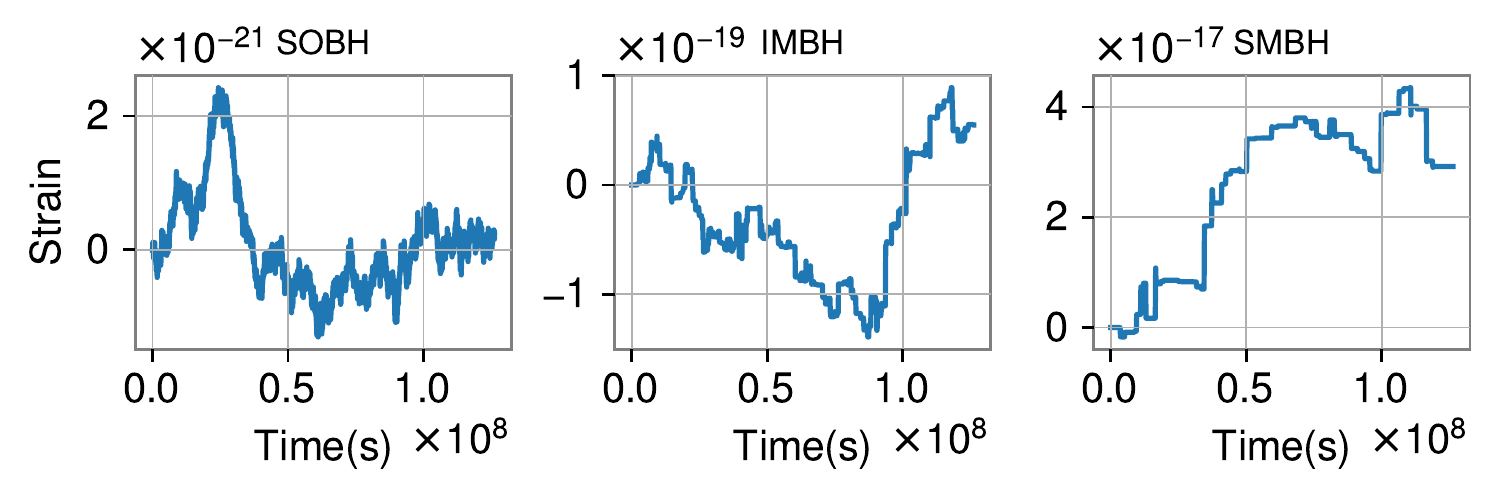}
    \caption{Three examples of the detector responses to SGWMB. Wave strain with respect to time is plotted. From left to right, the panels respectively show the SGWMB corresponding to SOBH, IMBH, and SMBH binaries.}
    \label{fig2}
\end{figure}

In Fig.~\ref{fig2} we show three typical examples of detector responses to the SGWMB. In the current work, we consider a detector with pattern functions
\begin{align}
F^+(\theta,\phi,\psi)&\equiv-\frac{1}{2}(1+\cos^2\theta)\cos2\phi\cos2\psi\nonumber\\
&\,\,\,\,-\cos\theta\sin2\phi\sin2\psi,\\
F^\times(\theta,\phi,\psi)&\equiv+\frac{1}{2}(1+\cos^2\theta)\cos2\phi\sin2\psi\nonumber\\
&\,\,\,\,-\cos\theta\sin2\phi\cos2\psi.
\end{align}
All kinds of detection schemes including pulsar timing array, space based interferometry and ground based interferometry admit similar pattern functions to the above one upto minor changement \cite{maggiore2008gravitational}.
Uniform distribution is applied to the parameters $\iota, \theta, \phi, \psi, \phi_c$. For the polar angles, $\cos\iota$ and $\cos\theta$ are uniformly distributed. Three types of BBH have been considered here including Stellar Original Black-hole (SOBH), Intermediate Mass Black-hole (IMBH), and SuperMassive Black-hole (SMBH), whose total mass range are respectively set as about $(10^1,10^2)M_\odot$, $(10^2,10^5)M_\odot$, and $(10^5,10^7)M_\odot$. The parameters $q, M, \vec{\chi}_1,\vec{\chi}_2, d_L$ are set according to the specific event rate of the binary merger events.

In Fig.~\ref{fig2}, a combination of M33.A and M43.A of Ref.~\cite{belczynski2020evolutionary,2021ApJ...921..156Z}, $\mu=0.01, \alpha=0.1, q=1$ case of Ref.~\cite{rasskazov2020binary} and LS-stalled model of Ref.~\cite{bonetti2019post} are used for respectively SOBH, IMBH and SMBH binaries. M33 and M43 are binary evolution models which can be used to deduce the event rate of stellar massive binary black holes \cite{2022LRR....25....1M}. The existence of intermediate-mass black holes is still elusive. Here we consider the formation of binary intermediate black holes in globular clusters and the corresponding event rate is adopted. The evolution of SMBH binaries may be driven by the stellar and/or gaseous environment and the third black hole. Different scenario results in different merger rates. Just as an example we use the driven mechanism through the third black hole to deduce the event rate in the right panel of Fig.~\ref{fig2}. Other scenarios will be discussed in the following sections.

As we mentioned before, SGWMB behaves as a sum of a lot of step functions with random amplitude and random sign. Such behavior is nothing but a one-dimensional Brownian motion $\mathfrak{M}(t)$. Apparently we have $\mathfrak{M}(0)=0$ and $\langle\mathfrak{M}(t)\rangle=0$ where $\langle\cdot\rangle$ means the ensemble average. Since $\mathfrak{M}(t_1)-\mathfrak{M}(0)$ and $\mathfrak{M}(t_2)-\mathfrak{M}(t_1)$ are independent $(t_1\neq t_2)$, we have
\begin{align}
&\langle[\mathfrak{M}(t_1)-\mathfrak{M}(0)][\mathfrak{M}(t_2)-\mathfrak{M}(t_1)]\rangle=\nonumber\\
&\langle\mathfrak{M}(t_1)-\mathfrak{M}(0)\rangle\langle\mathfrak{M}(t_2)-\mathfrak{M}(t_1)\rangle=0.
\end{align}
Consequently, we have $\langle\mathfrak{M}(t_1)\mathfrak{M}(t_2)\rangle=\langle\mathfrak{M}^2(t_1)\rangle$. The random amplitude and the random sign of the step functions are reduced from the independent random parameters of the binary mergers. Although the number of parameters is not huge, we can still assume the distribution of the step function is Gaussian \cite{PhysRevD.104.064056}. Consequently, we have
\begin{align}
\langle\mathfrak{M}^2(t)\rangle=2Dt,
\end{align}
where $D$ corresponds to the strength of the diffusion corresponding to the Brownian motion. We have checked the behavior of the Brownian motions shown in Fig.~\ref{fig1}. All of them satisfy the above relation which verifies our assumption. By numerical simulation, we get $D$ to be $3.16 \times 10^{-50}$, $8.42 \times 10^{-47}$, and $1.73 \times 10^{-42}$ for SOBH, IMBH, and SMBH respectively in Fig.~\ref{fig1}.

Let us assume the variance of each memory step function mentioned above is $\sigma$ and the average time between two consecutive memory step functions is $\Delta t$. We could write $D$ as \cite{krapf2018power}
\begin{align}
D=\frac{\sigma^2}{2\Delta t},\label{eq5}
\end{align}
which can be deduced from standard Brownian motion theory.

Combine the results of Ref.~\cite{liu2021accurate} and Ref.~\cite{PhysRevD.104.064056} the amplitude of the step function, which  corresponds to a single GW memory event, can be expressed as
\begin{align}
&\mathcal{A}=\frac{M}{D_L}F^+(\theta,\phi,\psi)Y_{-220}(\iota)[0.0969+0.0562\chi_{\rm up}+\nonumber\\
&0.0340\chi_{\rm up}^2+0.0296\chi_{\rm up}^3+0.0206\chi_{\rm up}^4](4\eta)^{1.65},\label{eq3}\\
&\chi_{\rm up}\equiv\chi_{\rm eff}+\frac{3}{8}\sqrt{1-4\eta}\chi_{\rm A},\\
&\chi_{\rm eff}\equiv(m_1\vec{\chi}_{1}+m_2\vec{\chi}_{2})\cdot\hat{N}/M,\\
&\chi_{\rm A}\equiv(m_1\vec{\chi}_{1}-m_2\vec{\chi}_{2})\cdot\hat{N}/M,
\end{align}
where $\hat{N}$ is the direction of the orbital angular momentum of the binary and $\eta\equiv\frac{m_1m_2}{M^2}$ is the symmetric mass ratio. Regarding SGWMB, parameters $m_{1,2}$, $\vec{\chi}_{1,2}$, $D_L$, $\iota$, $\theta$, $\phi$, and $\psi$ are random variables. Consequently, we have
\begin{align}
\sigma^2=\langle\mathcal{A}^2\rangle-\langle\mathcal{A}\rangle^2.
\end{align}
Since the parameters $m_{1,2}$, $\vec{\chi}_{1,2}$, $D_L$, $\iota$, $\theta$, $\phi$, and $\psi$ are independent of each other, we can treat them individually:
\begin{align}
&\mathcal{A}=\mathcal{A}_{\rm bbh}\mathcal{A}_{\rm ang},\\
&\mathcal{A}_{\rm bbh}\equiv\frac{M}{D_L}[0.0969+0.0562\chi_{\rm up}+\nonumber\\
&0.0340\chi_{\rm up}^2+0.0296\chi_{\rm up}^3+0.0206\chi_{\rm up}^4](4\eta)^{1.65},\\
&\mathcal{A}_{\rm ang}\equiv F^+(\theta,\phi,\psi)Y_{-220}(\iota).
\end{align}
The last equation is because the $h_{\times}$ part of GW memory in the source frame always vanishes \cite{liu2021accurate,PhysRevD.104.064056}.

Especially, the distributions of $\iota$, $\theta$, $\phi$, and $\psi$ are always uniform which makes $\langle\mathcal{A}_{\rm ang}\rangle=0$ and consequently $\langle\mathcal{A}\rangle=0$. After further calculation,
we get the corresponding variance as
\begin{align}
\langle\mathcal{A}_{\rm ang}^2\rangle-\langle\mathcal{A}_{\rm ang}\rangle^2\equiv\sigma^2_{\rm ang}=\frac{1}{20\pi}.
\end{align}

The distributions of $m_{1,2}$, $\vec{\chi}_{1,2}$, and $D_L$ are determined by the event rate of the binary systems. When the event rate information of the binary systems is given, we can calculate the rest part of $\sigma$
\begin{align}
&\sigma^2_{\rm bbh}\equiv\langle\mathcal{A}_{\rm bbh}^2\rangle-\langle\mathcal{A}_{\rm bbh}\rangle^2,\mu_{\rm bbh}\equiv\langle\mathcal{A}_{\rm bbh}\rangle,\\
&\sigma=\sigma_{\rm ang}\sqrt{\sigma^2_{\rm bbh}+\mu^2_{\rm bbh}}=\frac{1}{\sqrt{20\pi}}\sqrt{\sigma^2_{\rm bbh}+\mu^2_{\rm bbh}}.\label{eq4}
\end{align}

Corresponding to Fig.~\ref{fig2}, the above theoretical predictions (\ref{eq5}) and (\ref{eq4}) to the diffusion strengths of SOBH, IMBH, and SMBH are respectively $3.16\times10^{-50}$, $8.41\times10^{-47}$, and $1.73\times10^{-42}$. Our theoretical results are almost the same as the direct numerical simulation results which validate our theory for GW memory background.

Similar to the description of the usual stochastic gravitational wave background (SGWB), we investigate the power spectral density (PSD) of $\mathfrak{M}(t)$
\begin{align}
&S^\mathfrak{M}(f)\equiv\lim_{T\rightarrow\infty}\frac{1}{T} \left| \int_0^T e^{-2\pi i f t}\mathfrak{M}(t) dt \right|^2\\
&=\lim_{T\rightarrow\infty}\frac{1}{T}  \int_0^T \int_0^T dt_1 dt_2 \cos (2\pi f (t_1 -
    t_2))\langle\mathfrak{M}(t_1)\mathfrak{M}(t_2)\rangle\\
    &=\lim_{T\rightarrow\infty}\frac{D}{\pi^2f^2}  \left[ 1 - \frac{\sin (2\pi f T)}{2\pi fT} \right]\\
    &=\frac{D}{\pi^2f^2}.\label{eq2}
\end{align}

Equivalently we can also use the characteristic strain to describe the SGWMB \cite{maggiore2000gravitational}
\begin{align}
h^\mathfrak{M}_c(f)&=\sqrt{2fS^\mathfrak{M}}=\frac{1}{\pi}\sqrt{\frac{\sigma^2_{\rm bbh}+\mu^2_{\rm bbh}}{20\pi f\Delta t}},
\end{align}
from which we can see that the strength of SGWMB for binary black hole mergers depends on and only on the event rate. In order to compare the common convention used in the literature about stochastic gravitational wave background, we can also convert the above characteristic strain to the dimensionless energy density \cite{maggiore2000gravitational}
\begin{align}
\Omega^\mathfrak{M}_{\rm GW}(f)&=\frac{4\pi^2}{3H_0^2}f^3S^\mathfrak{M}(f)=\frac{2\pi^2}{3H_0^2}f^2(h^\mathfrak{M}_c(f))^2\\
&=\frac{f}{30\pi H_0^2}\frac{\sigma^2_{\rm bbh}+\mu^2_{\rm bbh}}{\Delta t},
\end{align}
where $H_0$ is the Hubble constant.

\section{Detectability of SGWMB}\label{sec:5}
The Brownian motion theory presented in the above section is a statistical description of SGWMB. In order to let such a statistical description valid, the detection time should be long enough. As for equilibrium statistics, the detection time should be longer than the ergodic time \cite{PhysRevE.67.041102,2004EL.....67..335L,2004PhyA..334..187C}. In the Brownian motion case, the observation time should be longer than the average time between two successive memory events. Intuitively, such a long enough observation time just makes the energy power of Brownian motion at every frequency bin happen. In another word, the averaged time scale between two successive memory events does not limit the valid range of frequency.

Besides the newly proposed SGWMB in the current work, the stochastic gravitational wave background \cite{maggiore2000gravitational,maggiore2018gravitational} may originated from compact binary coalescence \cite{PhysRevD.102.083501} and the cosmological perturbation in the early universe. Due to the unknown detail of the cosmological perturbation, the related power spectrum of the SGWB from the cosmological perturbation may be big or small for detection in the near future \cite{PhysRevLett.120.031301,PhysRevLett.122.201101,PhysRevLett.126.141303}. Compact binary coalescence may produce both non-memory SGWB and SGWMB. The dominant one between the non-memory SGWB and the SGWMB is easier to be detected.

\begin{figure}
\includegraphics[width=0.7\textwidth]{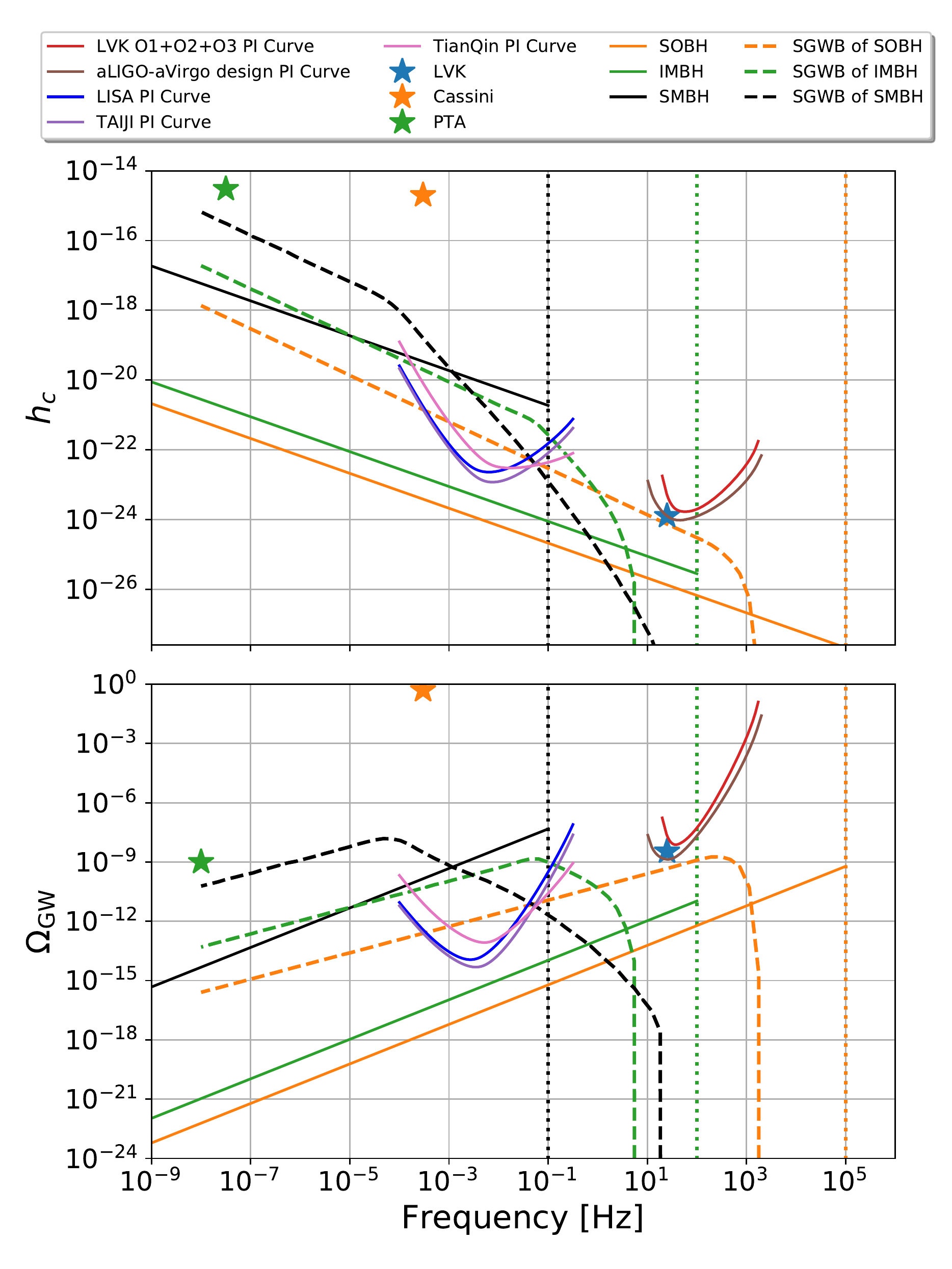}
    \caption{The characteristic strains (upper panel) and the dimensionless energy density (bottom panel) of SGWMB for several different typical event rates of BBH coalescence are compared. Together with the SGWMB, representative SGWB originated from the usual non-memory part of GW of BBH coalescence are shown as the dashed lines. The expected detection limits of LISA \cite{bavera2021stochastic}, Taiji \cite{PhysRevD.104.104015}, and Tianqin \cite{PhysRevD.105.022001} for SGWB based on their individual observation plan are shown as the blue, purple, and magenta lines respectively. These limits are quantified by the power-law integrated (PI) curve. The existing limits from LVK, Cassini, and pulsar timing array observations are shown as the stars. The PI curve for current ground-based detectors and their designed sensitivity are shown respectively as a red line and a brown line \cite{PhysRevD.104.022004}. The three vertical lines (red, green, and blue) correspond to the valid limit frequency of the SGWMB theory for different BBH classes (SOBH, IMBH, and SMBH).}
    \label{fig3}
\end{figure}

The power spectrum of the non-memory SGWB due to the compact binary coalescence can be divided into three parts which admit different power indexes \cite{zhu2011stochastic,PhysRevD.102.083501}. These three parts are highly related to the inspiral, merger, and ringdown stage of a BBH system. As shown in Fig.~\ref{fig3} the transition frequencies for SOBH, IMBH, and SMBH are respectively about thousands Hz, several Hz and mini Hz. Differently the power spectrum of the SGWMB always admits power index -2 as shown in (\ref{eq2}). So we can expect that the non-memory SGWB may dominate in some frequency regions, while the SGWMB may dominate in another frequency region. We plot our calculation results in Fig.~\ref{fig3}, from which we can see that the SGWMB dominates in about frequency region $f_{\rm merg}\lesssim f\lesssim\frac{0.5}{M}$. Here $f_{\rm merg}$ means the collection frequency between the inspiral part and the merger part of the non-memory SGWB and $M$ is the total mass of the typical BBH among the BBH coalescence ensemble. Corresponding to SOBH, IMBH and SMBH, such frequency ranges are respectively about $10^2{\rm Hz}<f<10^5{\rm Hz}$, $0.1{\rm Hz}<f<10^2{\rm Hz}$ and $10^{-4}{\rm Hz}<f<0.1{\rm Hz}$.

The strength of the stochastic background gravitational wave has been constrained by several experiments. Pulsar timing array (PTA) has limited the characteristic strain of the SGWB to less than about $h_c\lesssim3\times10^{-15}$ ($\Omega_{\rm GW}\lesssim10^{-9}$) at a reference frequency of about $10^{-8}$Hz \cite{lentati2015european}. Cassini data has been applied to limit the characteristic strain of the SGWB and constrain the amplitude to less than $h_c\lesssim2\times10^{-15}$  ($\Omega_{\rm GW}\lesssim0.5$) at a reference frequency of about 0.3mHz \cite{Armstrong_2003}. The O3a data of LIGO-Virgo-KAGRA Collaboration (LVK) can limit the characteristic strain of the SGWB to less than about $h_c\lesssim1.3\times10^{-24}$ ($\Omega_{\rm GW}\lesssim10^{-9}$) at reference frequency of about 25Hz \cite{PhysRevD.104.022004}. We plot these limit points in Fig.~\ref{fig3}.

According to the result shown in Fig.~\ref{fig3}, we do not expect that the pulsar timing array can detect SGWMB without a specially designed algorithm that can extract SGWMB buried in the total SGWB data. And the possibility for LVK to observe the SGWMB is also low. On the contrary, future space-based detectors including LISA, Taiji, and Tianqin may detect SGWMB of supermassive BBH because the SGWMB is stronger than the corresponding non-memory SGWB. In order to indicate such detectability, we plot the power-law integrated (PI) curve \cite{PhysRevD.104.022004,PhysRevD.88.124032} for LISA \cite{bavera2021stochastic}, Taiji \cite{PhysRevD.104.104015}, Tianqin \cite{PhysRevD.105.022001}, the current ground-based detectors and their designed sensitivity \cite{PhysRevD.104.022004} together with the kinds of non-memory SGWB in Fig.~\ref{fig3}. Possibly even the transition behavior of the power spectrum from non-memory background to memory background can be detected.

Besides the event rate for SMBH binary merger indicated in Fig.~\ref{fig3}, we have also tested several different event rates including all cases in Ref.~\cite{degraf2020morphological}, LS-Stalled and HS-Stalled cases in Ref.~\cite{bonetti2019post}, and all three cases in Ref.~\cite{klein2016science} whose diffusions are $D=2.53\times 10^{-44},~D=8.75\times 10^{-45},~{\rm and}~D=1.57\times 10^{-44}$ for popIII, Q3-delay, and Q3-nodelay models respectively \cite{yi2021gravitational}. LS-Stalled and HS-Stalled models describe SMBH binaries driven by the third black hole. Other than a simple model, detailed simulation involving galaxy formation and evolution, including primordial and metal-line cooling with a time-dependent UV background including self-shielding; star formation with associated supernova feedback; stellar evolution, gas recycling, and metal enrichment with mass and metal-loaded outflows, can be used to deduce the merger rate of SMBH binaries. The cases in Ref.~\cite{degraf2020morphological} fall in this category. The black hole seeding and the delays between the merger of two galaxies and the merger of the black holes hosted by those galaxies will strongly affect the merger rate of SMBH binaries. These two factors can be investigated through simulations. The cases in Ref.~\cite{klein2016science} fall in this category. For all of these event rates, the PI curve of LISA/Taiji/Tianqin is always much lower than the power spectrum of the corresponding SGWMB.

Quantitatively we can use the signal-to-noise ratio (SNR) to indicate the detectability \cite{Zhao:2020iew}
\begin{align}
&{\rm SNR}=\sqrt{T}\left[\int_0^\infty\frac{(S^\mathfrak{M})^2(f)}{S^2_{\rm n}(f)}df\right]^{1/2},
\end{align}
where $S_{\rm n}(f)$ is the detector sensitivity, and $T$ is the total observation time. The sensitivity for LISA/Taiji/Tianqin is listed in the Appendix. Respectively the SNR of SGWMB for LISA/Taiji/Tianqin is about $2\times10^{42}D\sqrt{T}$, $7\times10^{42}D\sqrt{T}$, and $8\times10^{41}D\sqrt{T}$. Roughly such SNRs for different event rate of SMBH merger are between 1 and thousands. So we can expect that future detection can distinguish these different event rates.

\section{Implication of SGWMB detection}\label{sec:6}
GW memory is an outstanding prediction of general relativity. But single GW memory event is extremely hard to detect \cite{Set09,VanLev10,PshBasPos10,CorJen12,MadCorCha14,Arzoumanian_2015,ZHANG2017743,PhysRevLett.117.061102,PhysRevLett.118.181103,PhysRevD.102.023010,PhysRevD.104.064056,2021arXiv211007754S}. The newly found SGWMB provides an alternative way to detect GW memory. From our Brownian motion model, we are sure that the power spectrum index of the SGWMB is always -2 which is independent of the gravity theory. Different gravity theory may result in different strength of SGWMB which can be tested through detection.

Similar to the usual non-memory SGWB, SGWMB also encodes the information of the tensor $+/\times$, vector $x/y$, and breathing/longitudinal polarizations \cite{Callister:2017ocg}. Methods designed for SGWB \cite{Chen:2021wdo} to extract that information could be applied to SGWMB straightforwardly. Such information can also be used to test gravity theory \cite{10.1093.nsr.nwx029,2021SCPMA..6420401B}.

The merger rate of supermassive BBH is highly uncertain. Especially, people are still debating what kind of mechanism drives the supermassive BBH to pass the final parsec. The PTA observation sensitivity gradually increases. But the corresponding SGWB detection of supermassive BBH is still missing. And more even PTA detected the corresponding SGWB, we are still not sure whether the BBH merges or not. This is because inspiral only binaries produce the dominant part of the non-memory SGWB. In contrast, the detection of SGWMB indicates the merger of BBHs without ambiguity.

Our calculation shows that space-based detectors are promising to detect SGWMB because such detector's frequency range is optimal for the SGWMB of supermassive BBH. So we expect LISA/Taiji/Tianqin are ideal for SGWMB detection. In the coming years, space-based detectors may realize the first detection of GW memory through SGWMB. Consequently, the last pc problem can be surely answered and the merger rate of supermassive BBH can be determined accordingly. On the contrary, if no SGWMB was found either general relativity or the theory of supermassive BBH merger should be checked carefully.

\vspace{2em}
\begin{acknowledgements}
We greatly thank Jin-Ping Zhu and He Wang for useful suggestions. We are also thankful to Xing-Jiang Zhu for his comments on the manuscript. This work was supported by the NSFC (No.~11690023, No.~11633001, No.~11920101003 and No.~12021003). Z. Cao was supported by ``the Interdiscipline Research Funds of Beijing Normal University" and CAS Project for Young Scientists in Basic Research YSBR-006.
\end{acknowledgements}

\bibliographystyle{unsrt}
\bibliography{refs}

\begin{appendix}
\renewcommand{\thesection}{Appendix}
\section{}
Regarding the sensitivity of space-based detectors, we use the following approximation (Eq.~(13) of \cite{robson2019construction})
\begin{align}
S_{\rm n}(f)&=\frac{10}{3L^2}\left(P_{\rm OMS}+2(1+\cos^2(f/f_*))\frac{P_{\rm acc}}{(2\pi f)^4}\right)\times\nonumber\\
&\left(1+\frac{6}{10}\left(\frac{f}{f_*}\right)^2\right),\\
f_*&=c/(2\pi L).
\end{align}
For LISA \cite{robson2019construction} we have
\begin{align}
P_{\rm OMS}&=(1.5\times10^{-11}{\rm m})^2{\rm Hz}^{-1},\\
P_{\rm acc}&=(3\times10^{-15}{\rm ms}^{-2})^2\left(1+\left(\frac{4\times10^{-4}{\rm Hz}}{f}\right)^2\right){\rm Hz}^{-1},\\
L&=2.5\times10^9{\rm m}.
\end{align}
For Taiji \cite{PhysRevD.102.024089} we have
\begin{align}
P_{\rm OMS}&=(8\times10^{-12}{\rm m})^2{\rm Hz}^{-1},\\
P_{\rm acc}&=(3\times10^{-15}{\rm ms}^{-2})^2\left(1+\left(\frac{4\times10^{-4}{\rm Hz}}{f}\right)^2\right){\rm Hz}^{-1},\\
L&=3\times10^9{\rm m}.
\end{align}
For Tianqin we have \cite{luo2016tianqin}
\begin{align}
P_{\rm OMS}&=(1\times10^{-12}{\rm m})^2{\rm Hz}^{-1},\\
P_{\rm acc}&=(1\times10^{-15}{\rm ms}^{-2})^2\left(1+\frac{1\times10^{-4}{\rm Hz}}{f}\right){\rm Hz}^{-1},\\
L&=\sqrt{3}\times10^8{\rm m}.
\end{align}

\end{appendix}

\end{document}